# Understanding User Perceptions of Human-centered AI-Enhanced Support Group Formation in Online Healthcare Communities

Pronob Kumar Barman, MS[1], James R. Foulds, PhD[1], Tera L. Reynolds, PhD, MPH, MA[1]
[1]University of Maryland, Baltimore County, Baltimore, Maryland, USA

**Abstract**
*Peer support is critical to managing chronic health conditions. Online health communities (OHCs) enable patients and caregivers to connect with similar others, yet their large scale makes it challenging to find the most relevant peers and content. This study assessed perceived value, preferred features, and acceptance conditions for algorithmically personalized support group formation within OHCs. A two-phase, mixed-methods survey (N=165) examined OHC participation patterns, personalization priorities, and acceptance of a simulated personalized support group. Perceived value of the simulated support group was high (mean 4.55/5; 62.8% rated 5/5) and 91.5% would join this group. The importance participants placed on peer matching strongly correlated with perceived value ($\rho=0.764$, $p<0.001$). Qualitative findings revealed conditional acceptance: participants demand security, transparency, human oversight, and user control over data. Personalized support groups may be desired, but they will not be adopted unless trust, privacy, and algorithmic governance concerns are addressed.*

**Introduction**

Living with chronic health conditions involves complex challenges extending beyond clinical treatment—emotional adjustment, symptom management, lifestyle adaptation, and navigating healthcare systems.[1] Patients and caregivers often need support beyond what infrequent clinical encounters can provide, particularly experiential knowledge and emotional validation from others with similar health journeys.[1,4] Online health communities (OHCs) have emerged as vital platforms for this purpose, enabling individuals to share lived experiences, access informational and emotional resources, and reduce feelings of isolation.[2,3,4,5] Social support theory posits that peers fulfill distinct functions—informational, emotional, and appraisal—and that perceived similarity in diagnosis, treatment stage, and psychosocial needs strengthens the credibility and relevance of support received.[9,10,11,12,13]

However, as OHCs grow in scale, their broad-scope forums present challenges. With millions of users, discussion feeds become overwhelming, creating information overload and making it difficult to identify peers whose experiences match one's own health circumstances or psychosocial needs.[5,41] While search and filtering tools help users locate relevant posts, they do not address a more fundamental need: sustained, reciprocal relationships with well-matched peers who provide ongoing support. Effective peer support requires forming connections with individuals who share similar conditions, goals, and communication preferences—something passive content discovery cannot achieve.[4,5,13] Smaller, algorithmically personalized support groups offer a potential solution by proactively connecting users with compatible peers, moving beyond reactive information retrieval toward active community formation.

Recent work has also attempted to leverage algorithmic approaches for recommending support groups to OHC users, though achieving accurate and personalized group matching remains an open challenge.[13] Existing research has explored both the technical feasibility of matching users by clinical and psychosocial similarity[33,34,35] and patient perceptions of algorithms in consumer health technologies, finding that attitudes are highly context-dependent—shaped by privacy, algorithmic trust, fairness, and control concerns.[24,26,43,44] However, limited human-centered research examines how users perceive algorithmic peer matching for support group formation in OHCs—a context involving sensitive health disclosures, ongoing relational dynamics, and higher stakes for mismatched groupings—and what design features, privacy safeguards, and governance structures would make such systems acceptable.[34,35,45] We address this gap with a two-phase mixed-methods study integrating quantitative preference measures with qualitative accounts of conditional acceptance.

This study addresses three overarching research questions:

*RQ1:* How do patients weigh the perceived benefits and drawbacks of personalized support groups compared to non-personalized OHCs?

*RQ2:* Which features, and design elements of personalized support group systems would enhance satisfaction, communication, and intention for future engagement?

*RQ3:* How do perceived privacy risks, trust in algorithmic group assignment, and willingness to share personal data affect acceptance of personalized support group systems?

**Methods**

We conducted a two-phase mixed-methods online survey. Phase 1 captured OHC participation and baseline preferences. Approximately one week later, we simulated how a support group generation algorithm—building on our prior work[13]—is intended to work to manually create a fictional support group based on Phase 1 preferences. Phase 2 presented this "algorithm"-generated group and collected evaluations of willingness to join, satisfaction, and design priorities. This study was approved as exempt survey research (IRB Protocol #1564; Exempt Category 2). Figure 1 provides an overview.

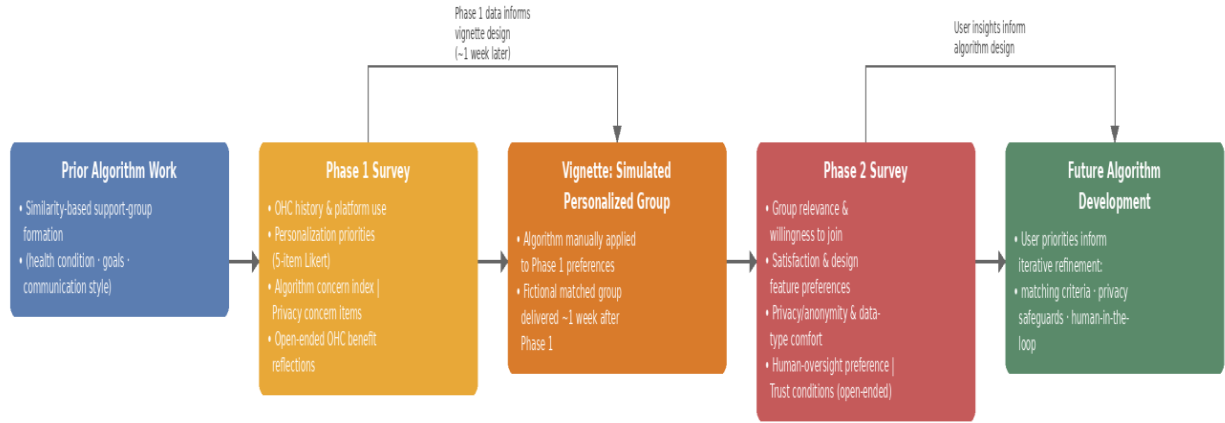

**Figure 1.** Overview of the two-phase mixed-methods survey design and analysis.

*Survey Development*

We used an iterative approach to develop the two online survey instruments in Qualtrics.[46] We initially drafted survey questions based on our research questions, existing theory, and our prior work to develop an algorithm to form support groups. We then conducted a pilot and used those results to refine and finalize the instruments.

Initial Survey Instrument Development

Our survey instruments were informed by three theoretical frameworks—social support and homophily, privacy calculus and data disclosure, and technology acceptance models—and the literature on human-centered AI and algorithmic trust. Relevant constructs from each framework (perceived similarity, privacy concern dimensions, perceived usefulness, trust in algorithms, willingness to share health data) formed question groupings across both phases.

*Social support and homophily.* Homophily—the tendency for similar individuals to form connections[14]—motivated our personalization and peer-matching measures. Perceived similarity in health experiences enhances engagement and support value,[9,10,11,12,13] informing our Phase 1 Likert items on matching dimensions and Phase 2 satisfaction measures.

*Privacy calculus and data disclosure.* Because personalized matching requires sensitive health data, we drew on privacy calculus models—where individuals weigh anticipated benefits against perceived privacy risks[15,16]—to develop privacy concern and data-sharing items. Disclosure decisions depend on institutional safeguards, governance transparency, and user control,[7,17,18,19,20,21] guiding our inclusion of select-all-that-apply privacy concern items (Phase 1) and open-ended questions about data-sharing conditions (Phase 1).

*Technology acceptance and facilitating conditions.* The Technology Acceptance Model (TAM)[30] and the Unified Theory of Acceptance and Use of Technology (UTAUT)[31,32] suggest that perceived usefulness, ease of use, and facilitating conditions (e.g., privacy controls, governance structures) shape behavioral intentions to adopt new systems.[30,31,32] These constructs informed our measures of perceived added value and willingness to join (Phase 2), interest in group activities (5-point Likert scale), feature priority rankings (Phase 2), and satisfaction ratings (Phase 2).

*Human-centered AI and algorithmic trust.* Algorithm concern measures were informed by human-AI interaction research revealing context-dependent preferences: algorithm aversion after observing errors or overreliance without appropriate skepticism.[22,23,24,25] In healthcare, trust is shaped by transparency, accountability, and human oversight mechanisms,[8,26,27] and explainability interventions can paradoxically increase overreliance, requiring careful design of transparency features.[28] Fairness concerns arise when algorithms differentially impact who is grouped, matched, or excluded.[29] These motivated our algorithm concern index (Phase 1), trust condition questions (Phase 1), and inclusion of human oversight in the Phase 2 scenario.

Phase 1 also included demographics (age, gender, race/ethnicity, country), health context (chronic conditions count; current health need), OHC participation type and frequency, perceived added value of personalization (5-point Likert), personalization priorities (five 5-point Likert items covering health condition, treatment stage, communication preferences, and health goals), and select-all-that-apply concerns about algorithm use and privacy/data-sharing.

Pilot and Survey Instrument Refinement

We conducted a pilot study with five participants (two Black [one male, age 28; one female, age 35], two White [two female, ages 30 and 44], and one Asian [one male, age 30]) recruited from the CARE and Latent research labs within the UMBC Information Systems department. Pilot participants confirmed that survey language and item wording were clear. The Phase 1 survey took an average of 17 minutes (Range: 15–20 minutes). However, one participant noted that the vignette review required substantial critical thinking and back-and-forth reflection, resulting in a completion time of approximately 25–30 minutes rather than the originally estimated 5–10 minutes. Based on this feedback, we revised the estimated completion time for Phase 2 to approximately 25–30 minutes (total across both sessions: 50–60 minutes). The pilot also confirmed the feasibility of the two-phase design.

*Vignette Development*

We used a vignette-based evaluation approach to understand people's opinions on a simulated personalized support group.[36] This approach enabled evaluation of trust, privacy, and design requirements, which are critical to understand before building the actual system. The vignette was developed to simulate the output of the group formation algorithm (gDMR; Group-specific Dirichlet Multinomial Regression and gSTM; Group-specific Structural Topic Model), which matches users based on user-generated content, demographic features (age, gender, geographic location), and platform engagement data (membership duration, included as a proxy for familiarity with community norms and depth of engagement).[13] The matching objective was to form groups of individuals as similar as possible across these dimensions. Each vignette comprised four components: (1) an Assigned Participant Profile representing the respondent's own matching attributes and rationale (homophily matching priority: health condition → demographic features [age, gender, geographic location] → membership duration); (2) a Group Profile with standard features, including privacy features, size of group, and communication modality (e.g., pseudonymous usernames, 10–15 members, text-based); (3) an Example Group Members table (four matched peers); and (4) a Group Activities menu (Curated Thread Highlights, Interactive Polls, Viewer's Digest). Participants reviewed their vignette about 1 week after Phase 1, then completed Phase 2.

*Participant Recruitment*

Recruitment occurred online via health-focused social media platform posts (including Facebook health groups and Reddit health sub-reddits) and department email lists distributed to undergraduate and graduate students at UMBC. Eligible participants were adults (18+) who reported at least monthly engagement in an online health forum or support group in the last 12 months. Participation was voluntary; respondents who completed both phases of the study were eligible for a raffle draw for a $20 gift card (up to 35 winners).

*Data Analysis*

We computed descriptive statistics for all variables. For RQ1, we examined willingness and perceived value distributions and Spearman correlations. Planned chi-square tests of willingness by participation type were infeasible due to sparse cells (expected counts <5); we report descriptive patterns instead. For RQ2, we summarized activity interest and feature ranks with subgroup comparisons (Mann–Whitney U, Kruskal–Wallis) and Benjamini–Hochberg FDR correction.[37] For RQ3 (privacy and algorithmic trust), we assessed distributions of the algorithm concern index (inverse count of algorithm concerns; range 0–5) and the privacy concern count (count of privacy concerns; range 1–4), then compared these indices between Yes versus Maybe respondents using Mann–Whitney U tests; because no participant declined outright (0 'No' responses), this contrast represented the only available indicator of conditional acceptance.

Open-ended responses were analyzed via reflexive inductive thematic analysis.[38] Consistent with reflexive thematic analysis, the first author (PKB) served as sole coder—an approach Braun and Clarke consider philosophically coherent. To establish rigor, PKB engaged in weekly peer debriefing with two faculty advisors (T.L.R. and J.F.), who provided iterative feedback on codebook development and emerging themes, serving as an ongoing audit trail. Coding was conducted in a spreadsheet with documented codebooks for each research question. PKB reviewed all English responses; non-English responses were excluded (n=28, 7%). Codes were grouped into higher-order categories; themes were derived by reviewing for overarching interpretive patterns and refined until thematic saturation— operationalized as no new codes emerging from additional responses. Quantitative and qualitative findings were integrated using joint displays to generate meta-inferences.[39]

**Results**

A total of 165 respondents completed both phases of the study and comprise the analytic sample. Because we do not know how many people saw our recruitment advertisements on social media, we are unable to calculate a response rate. Demographically, the analytic sample (N=165) averaged 31.4 years of age (SD 5.7, N=161, range 20–49), were 44.8% female (n=74), and resided primarily in the United States (64.2%, n=106) and Viet Nam (28.5%, n=47). Participants reported living with a mean of 5.23 chronic conditions (SD 2.81); the three most commonly reported health concern types were mental health conditions (67.3%, n=111), caregiving support (65.5%, n=108), and chronic illness (54.5%, n=90). 53.3% (n=88) reported a health-related need at baseline (e.g., active symptom management, ongoing treatment, or emotional or informational support). Most respondents (69.7%, n=115) used both forums and support groups, with modal engagement at weekly or more often. Table 1 summarizes key participant characteristics.

**Table 1.** Participant characteristics (N=165). Reference column from [47] and [48]. NR = Not reported.

| **Characteristic** | **Sample** | **Online Health User Population (Reference)** |
|---|---|---|
| Age (years), mean (SD) | 31.4 (5.7) | ~43 (modal group: 45–64)[47] |
| Gender<br>    Man<br>    Woman<br>    Other or Prefer not to say | <br>53.9%<br>44.8%<br>1.2% | <br>Unknown<br>~56-78%[47,48]<br>Unknown |
| Country<br>    USA<br>    Viet Nam<br>    Other | <br>64.2%<br>28.5%<br>7.3% | |
| Race/ethnicity<br>    White<br>    Asian<br>    Black or African American<br>    Hispanic<br>    Other | <br>35.8%<br>32.1%<br>26.1%<br>1 (0.6%)<br>6.1% | <br>64.8%[48]<br>Unknown<br>8.9%[48]<br>3.2%[48]<br>- |
| Chronic conditions count, mean (SD) | 5.23 (2.81) | Unknown |
| Health-related need at baseline (e.g., active symptom management)<br>    Yes<br>    No<br>    Prefer not to say | <br><br>53.3%<br>43.6%<br>2.4% | <br><br>Unknown |
| OHC participation type<br>    Health Forum Only<br>    Support Group<br>    Both | <br>26.1%<br>4.2%<br>69.7% | <br>Unknown<br>~7%[48]<br>Unknown |
| [†]Population estimates are based on U.S. population-based surveys. | | |

*RQ1: Benefits and Drawbacks of Non-Personalized versus Personalized OHCs*

Inductive thematic analysis of 125 open-ended responses (79% substantive response rate) yielded six benefit themes from non-personalized OHC experiences: community and belonging (74% of coded responses), informational support (66%), applied learning and behavior change (49%), emotional support and validation (46%), helping others (11%), and psychological safety (2%). No drawback themes emerged despite the survey explicitly soliciting them.

In the community and belonging theme, P038 wrote: *"Reduced isolation—feeling less alone in your struggles."* An illustrative informational support example was, *"Sharing knowledge and experiences from others has shaped my*

*perspective on health" (P073).* In the applied learning and behavior change theme, participants described translating peer-shared strategies into daily health management practices.

Nonjudgmental environments were valued: *"A safe space—allows discussion of embarrassing symptoms, fears and frustrations without judgement*" (P140).

While participants valued non-personalized communities, they also perceived added value from personalization. In Phase 1, prior to viewing the simulated group, perceived added value ratings were high (Q31; mean=4.55/5, SD=0.68; 62.8% rated 5/5), assessing anticipated value in the abstract without direct comparison to non-personalized groups. Following Phase 2 vignette exposure, willingness to join was high (91.5% "Yes"; 8.5% "Maybe, with modifications"), with elevated satisfaction (peer-matching: mean=4.70/5; overall: mean=4.68/5). Phase 2 group relevance ratings (mean=4.75/5, SD=0.48) significantly exceeded pre-vignette expectations (Wilcoxon signed-rank: p=0.0007), suggesting vignette exposure strengthened perceived value. Additional outcomes confirmed acceptance: 89.5% reported the group would meet health needs (mean=4.65/5), mean engagement likelihood was 4.70/5, and 92.7% (n=153) would recommend the group. Figure 2 summarizes perceived added value and willingness to join.

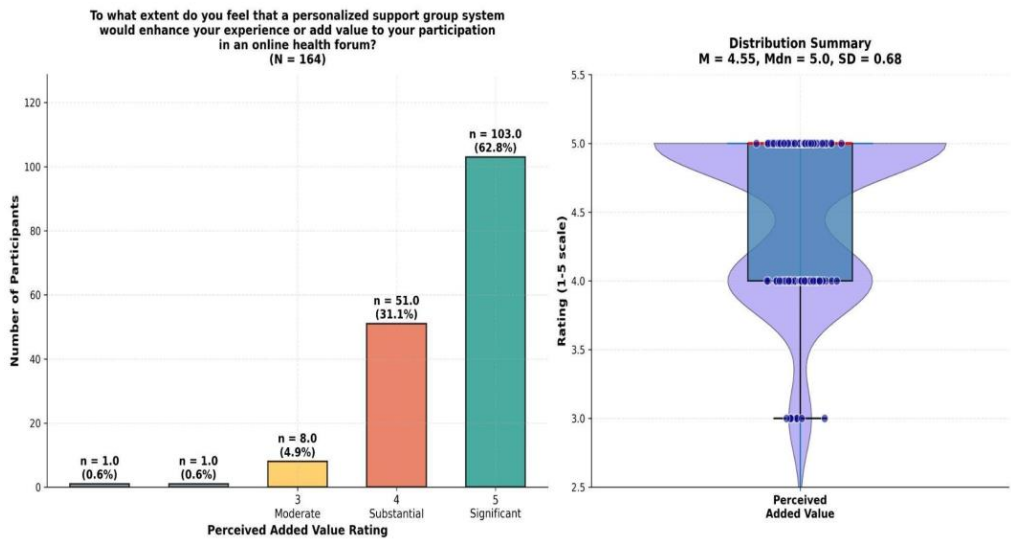

**Figure 2.** Perceived added value rating distribution (1–5 scale).

In Phase 1, ratings of personalization importance (Q40; five aspects rated 1–5: peer matching based on health data and social preferences, tailored content specific to health condition, personalized notifications, personalized participation suggestions, and customizable communication settings) were positively associated with perceived added value and Phase 2 satisfaction outcomes. Peer-matching importance showed the strongest association with perceived added value ($\rho$=0.764, p<0.001).

*RQ2: Desired Features and Activities for Personalized Support Groups*

In Phase 2, participants reported high interest across three activity families (3-option scale recoded: Definitely=5, Maybe=3, Not at all=1): goal management (mean 4.61, SD 0.88), knowledge and resource exchange (mean 4.59, SD 0.95), and synchronous engagement (mean 4.55, SD 0.87). In feature rankings (1=most important, 10=least; lower=higher priority), secure and private communication emerged as the top priority (mean rank 2.36, SD 1.81), followed by personalized recommendations (mean rank 2.95, SD 1.79). Lowest priorities were health tracking integration (mean rank 7.34, SD 1.69) and video/audio chat (mean rank 7.61, SD 1.95). Figure 3 summarizes activity interest by family.

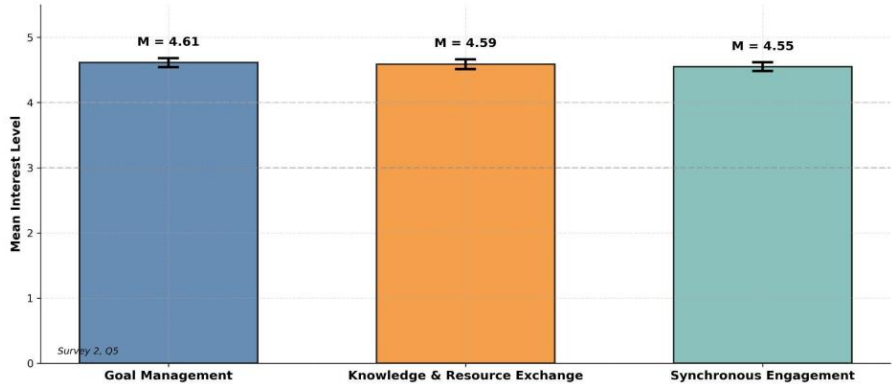

**Figure 3.** Phase 2 interest in support-group activities across three families (N=165; all three activity options shown). Scale: Definitely Interested (recoded=5), Maybe Interested (recoded=3), Not at all Interested (recoded=1).

Subgroup comparisons (FDR-corrected) revealed significant feature ranking differences. Participants with more conditions assigned higher importance to peer support ($p_{adj}<0.001$), secure communication ($p_{adj}=0.0024$), and regular notifications ($p_{adj}=0.009$). Those with active health-related needs similarly prioritized peer support ($p_{adj}=0.0017$) and secure communication ($p_{adj}=0.0110$). Female respondents prioritized tracking app integration ($p_{adj}=0.0110$). No significant differences in feature rankings were found by age (all $\rho<0.15$, all $p>0.05$ after FDR correction).

Qualitative findings nuance these rankings. Despite synchronous engagement ranking lower in feature importance, participants valued live interaction for emotional depth. However, participants managing multiple conditions found daily sharing burdensome, highlighting a design tension between synchronous connection and low-burden participation necessary for sustained engagement.

### RQ3: Privacy Risks, Algorithm Concerns, and Acceptance

The algorithm concern index (inverse count, 0–5; higher=fewer concerns) averaged 3.34 (SD 1.59, N=164); privacy concern count (1–4) averaged 2.43 (SD 1.08, N=156). Neither differed significantly by willingness to join (algorithm: p=0.449; privacy: p=0.543), likely reflecting the skewed willingness distribution (91.5% "Yes") limiting statistical power. Figure 4 displays distributions. In Phase 2, comfort with privacy/anonymity features was high (mean=4.70/5, SD=0.53; 73.3% rated 5/5), as was comfort with algorithmic data use across all dimensions (means 4.54–4.61). Furthermore, 89.1% (n=147) preferred human involvement in matching, underscoring desire for hybrid human–algorithmic oversight.

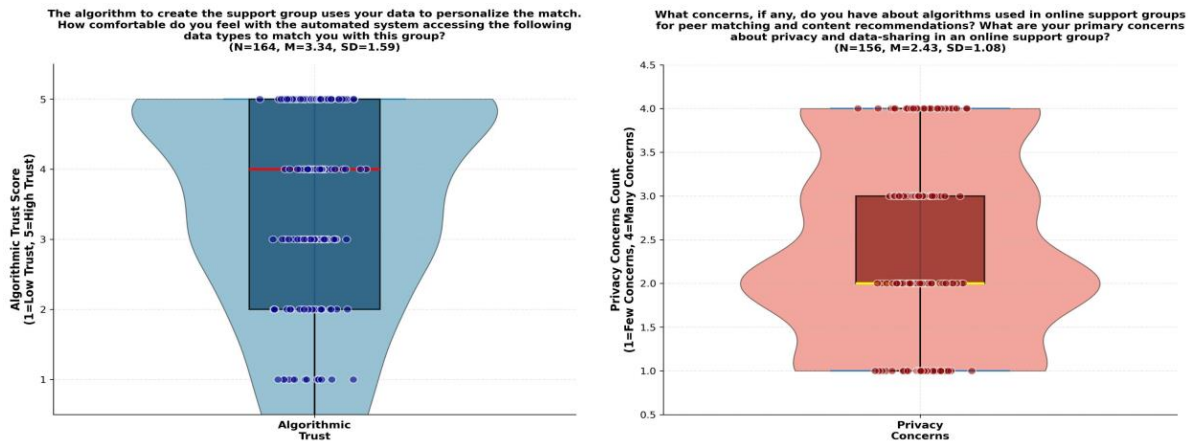

**Figure 4.** Distributions of algorithm concerns and privacy concerns (Phase 1). Left: algorithm concern index (0–5; higher = fewer concerns; N=164, M=3.34, SD=1.59). Right: privacy concerns count (1–4; N=156, M=2.43, SD=1.08).

Qualitative findings reveal conditional acceptance. Data security fears dominated privacy concerns (34% of 224 coded responses): *"I fear my information might be hacked or leaked in a data breach"* (P044, willingness-to-join: yes). Participants demanded consent, control, and transparency (69% of responses): *"I'm worried the system might not match me with people who really fit my preferences, and I don't feel like I have much control or understanding of how it makes those choices"* (P055, willingness-to-join: yes; algorithm concern index=3, indicating moderate concern).

Human oversight over purely algorithmic decision-making emerged as critical (20% of responses). Even willing participants expressed reservations: *"I worry that the system could feel impersonal compared to human-led group formation"* (P050, willingness-to-join: yes; algorithm concern index=1). Pragmatic data sharing (16%) reflected conditional acceptance contingent on demonstrable benefits: *"I think it's acceptable for the algorithm to use my personal data if it truly improves group matching and provides more relevant support. As long as my privacy is protected, I don't mind because the benefit of connecting with people who share similar needs outweighs the concern"* (P026, willingness-to-join: yes; algorithm concern index=1).

These findings illustrate a nuanced acceptance calculus: high stated willingness coexists with specific conditions—security, transparency, human oversight, and user control—required for actual adoption.[6,8,27]

**Discussion**
Three principal findings emerged. First, participants who placed higher importance on peer matching reported higher perceived value ($\rho=0.764$, $p<0.001$), and nearly all (91.5%) expressed willingness to join—indicating algorithmically matched groups are both desired and credible. Second, security and personalization are foundational: users ranked secure communication (mean rank 2.36) and personalized recommendations (mean rank 2.95) highest. Qualitative data highlighted a design tension between valued synchronous connection and low-burden participation for sustained engagement. Third, high stated willingness coexisted with conditional acceptance: 69% of open-ended responses articulated specific requirements for consent, transparency, human oversight, and user control, revealing that willingness alone does not guarantee uptake.

**Comparison With Prior Work and Existing Theory**
Our findings extend OHC research[2,3,4,41] and health recommender systems work[33,34,35] by showing that technical feasibility alone does not ensure willingness; human-centered design must bridge capability and uptake. Similar to Longoni et al.[24] on medical AI resistance, our participants expressed discomfort with fully automated decision-making; however, peer-matching appears to generate less resistance than clinical AI, possibly because perceived stakes differ. Esmaeilzadeh[43] found that transparency and data control were top concerns for consumer health AI tools—our results confirm these as equally central in the OHC peer-matching context. The conditional acceptance patterns align with prior work on privacy concerns and platform governance expectations.[6,7,17]

Our findings align with social support theory and homophily, which theorize that perceived similarity enhances support credibility. While this study did not directly measure trust in peer advice, high satisfaction and willingness-to-join ratings suggest participants perceived the matched group as credible and relevant.[9,10,11,13,14,40] Conditional acceptance patterns align with privacy calculus models.[15,16,18,19,20,21] We observe mixed algorithmic appreciation and aversion: high willingness coexists with demands for explainability, contestability, and human oversight.[8,22,23,24,25,26,27,28] Acceptance dynamics map onto technology acceptance models where perceived usefulness, control, and facilitating conditions shape behavioral intentions.[30,31,32]

**Implications for Human-Centered AI Design and Policy**
Our findings inform design requirements for trustworthy algorithmic peer matching. Our results suggest that platforms be transparent and contestable by having explainable matching criteria and the ability to override the groups that are formed.[8,27,28] Platforms must enable granular privacy controls, allowing users to specify which data types inform matching versus remain private.[15,17,18,19,20] Data breach fears were universal; security infrastructure and credible safeguard communication are non-negotiable. Concerns about the perceived limitations of algorithms in understanding the emotional and contextual nuances of health experiences suggest hybrid models combining algorithmic efficiency with human oversight and escalation pathways.[6,8,27,42] Finally, regular fairness audits of matching outcomes across demographic groups and accessible redress mechanisms are recommended to prevent inequitable groupings.[29]

**Limitations and Future Work**
This study used a vignette-based hypothetical scenario[36] rather than a deployed system, introducing potential gaps between stated and actual behavior. Self-selection likely favored participants with positive OHC experiences, and the skewed willingness distribution (91.5% "Yes"; 8.5% "Maybe"; 0% "No") precluded multivariable modeling and limited chi-square tests due to small expected cell counts. The RQ1 open-ended prompt yielded exclusively positive reflections, precluding characterization of OHC drawbacks—a gap given documented misinformation, conflict, and overload in online communities. Vietnamese-language responses (n=28, 7% of qualitative data) were excluded due to resource constraints, potentially missing culturally specific concerns. Social media recruitment precluded response rate calculation, limiting nonresponse bias assessment. Some survey fields exhibited missingness and ranked privacy items had formatting inconsistencies, constraining certain analyses. The sample's composition—predominantly younger adults (mean age 31.4), overrepresenting U.S. (64.2%) and Vietnamese (28.5%) participants with high prior OHC engagement—raises representativeness concerns. High willingness rates and positive perceived value scores may reflect pre-existing positive orientations toward technology rather than attitudes of the broader chronic illness population. Findings should be interpreted as reflecting engaged, technology-comfortable OHC users and may not generalize to older adults, less digitally engaged users, or non-Western communities with different platform norms and privacy expectations.

Future research should evaluate real-world deployments—including engagement, support quality, well-being, fairness, and trust construct validation. Implementation studies can test governance models and examine how preferences evolve with system experience.

**Conclusion**

Personalized algorithmic support group formation in OHCs is both desired and contested. In our two-phase mixed-methods study, nearly all participants expressed willingness to join algorithmically matched groups, with those rating peer matching as more important reporting higher perceived value. However, qualitative findings reveal this willingness is conditional: participants articulated requirements for data security, transparency, human oversight, and user control. Acceptance is not binary but negotiated—participants will engage with algorithmic matching *if* human-centered design principles are embedded in governance. These findings inform design requirements: transparent and contestable matching criteria, granular privacy controls, robust security, and human-in-the-loop oversight.

**Acknowledgments**
We are grateful to our participants for their time and for the conceptual input of Drs. Shimei Pan, PhD and Nirmalya Roy, PhD at the University of Maryland, Baltimore County.